\documentclass[%
 prb,
 amsmath,amssymb,
reprint,%
superscriptaddress,
floatfix,
]{revtex4-2}

\usepackage{graphicx,float}
\usepackage{bm,braket}
\usepackage{dsfont}
\usepackage[utf8]{inputenc}
\usepackage[T1]{fontenc}
\usepackage{mathptmx}
\usepackage{tikz}
\usetikzlibrary{positioning}
\usepackage{lipsum}
\usepackage{dsfont,siunitx}
\usepackage{hyperref}
\usepackage{soul}
\usepackage{amsmath}
\begin{document}


\title[Draft]{Efficient simulation of open quantum systems coupled to a reservoir through  multiple  channels}

\author{Hanggai Nuomin}%
\thanks{These authors contributed equally.}

\affiliation{
Department of Chemistry, Duke University, Durham, North Carolina 27708, United States
}%
\author{Jiaxi Wu}%
\thanks{These authors contributed equally.}
\affiliation{
Kuang Yaming Honors School, Nanjing University, Nanjing, Jiangsu 210023, China
}
\affiliation{ 
Department of Physics, Duke University, Durham, North Carolina 27708, United States
}%
\author{Peng Zhang}%
\email{peng.zhang@duke.edu.}
\affiliation{ 
	Department of Chemistry, Duke University, Durham, North Carolina 27708, United States
}
\author{David N. Beratan}%
\email{david.beratan@duke.edu.}
\affiliation{ 
	Department of Chemistry, Duke University,
	Durham, North Carolina 27708, United States
}
\affiliation{ 
	Department of Physics, Duke University,
	Durham, North Carolina 27708, United States
}
\affiliation{ 
	Department of
	Biochemistry, Duke University,
	Durham, North Carolina 27710, United States
}


\date{\today}

\begin{abstract}
The simulation of open quantum systems coupled to a reservoir through multiple channels remains a substantial challenge. This kind of open quantum system arises when considering the radiationless decay of excited states that are coupled to molecular vibrations, for example. We use the chain mapping strategy in the interaction picture to study systems linearly coupled to a harmonic bath through multiple interaction channels. In the interaction picture, the bare bath Hamiltonian is removed by a unitary transformation (the system-bath interactions remain), and 
a chain mapping transforms the bath modes to a new basis. The transformed Hamiltonian contains time-dependent local system-bath couplings. The open quantum system is coupled to a limited number of (transformed) bath modes in the new basis.  As such, the entanglement generated by the system-bath interactions is local, making efficient dynamical simulations possible with  matrix product states. We use this approach to simulate singlet fission, using a generalized spin-boson Hamiltonian. The electronic states are coupled to a vibrational bath both diagonally and off-diagonally. This approach generalizes the chain mapping scheme to the case of multi-channel system-bath couplings, enabling the efficient simulation of this class of open quantum systems using matrix product states.
\end{abstract}

\maketitle

\section{Introduction}
Dynamical simulations of open quantum systems have been significantly advanced by the development of chain-mapped time-dependent density-matrix renormalization group (DMRG) methods.  These methods enhance the efficiency of numerical simulations  \cite{white1993density,white1992density,white2004real,xu2022stochastic,liu2022suppressing,ren2022time,chin2010exact,prior2010efficient}. A bath continuum is mapped to a semi-infinite chain of modes (e.g., bosons or fermions) with nearest-neighbor couplings using an orthogonal transformation in the chain mapping method. As a result, the open quantum system in the transformed basis is coupled only to the first mode of the chain. It was found that the chain Hamiltonian has significantly slower entanglement growth during its time evolution than the untransformed Hamiltonian  \cite{liu2021improved}. In the transformed basis, the quantum state of the system and bath lives in a small Hilbert space spanned by a few dozen important states (adaptively generated by the DMRG algorithm). As a consequence, the chain representation is particularly well suited for efficient DMRG simulations. However, chain mapping is only possible when a single spectral density describes the open quantum system. That is, for a system interacting with a bath through many kinds of system-bath couplings (i.e., multi-channel couplings  \cite{broadbent2012solving}), the mapping of the bath to a nearest-neighbor  chain is not feasible. Efficiently simulating open quantum systems of this class  using DMRG, or equivalently, using matrix product states (MPS), remains an open challenge  \cite{nusseler2020efficient};  however, open quantum systems with multi-channel couplings are ubiquitous in chemical dynamics. Bridge-mediated electron transfer in donor-bridge-acceptor complexes   \cite{li2020symmetry,lin2009modulating,skourtis1993two,beratan2015charge} provides a familiar example of such open quantum systems, as the donor and acceptor electronic energies and couplings are influenced by nuclear motions. For this reason, developing efficient algorithms to simulate the time evolution of multi-channel open quantum systems is particularly timely.

In earlier studies  \cite{liu2021improved}, the mapped chain Hamiltonian was transformed into the interaction picture with respect to the reservoir (bath) Hamiltonian, referred to as the interaction picture chain (IC) Hamiltonian. In the IC Hamiltonian, the system-bath coupling coefficients are time-dependent, in contrast to the constant coefficients before transformation (i.e., the Schr\"odinger picture). The IC time-dependent coupling coefficients are found to be localized, and the system-bath interactions reside in a narrow window of bath modes and have a travelling wave pattern  \cite{polyakov2022real} that resembles  a chain structure. That is, the system interacts with a small number of bath modes at all times, enabling efficient simulations.

Here, we apply the interaction picture chain mapping to multi-channel open quantum systems and study the pattern of time-dependent system-bath couplings. As an example, we simulate the population dynamics and entanglement growth of a singlet fission process  \cite{HuangSFmodel} mediated by two types (diagonal and off-diagonal) of vibronic couplings (two distinct spectral densities and one bath), as an example. The singlet fission system was  studied earlier by Zhao et al. using the multiple Davydov ansatz  \cite{zhao2022hierarchy,shen2022simulation}. Our simulations show slightly different dynamics from theirs, while both methods are exact in principle, given sufficiently tight convergence thresholds. The disagreement in simulation results requires a careful reexamination of the system studied in Ref~\cite{zhao2022hierarchy}. 

Our method extends the chain-mapped time dependent DMRG in the interaction picture to the multi-channel open quantum systems. The efficiency of the method is ensured by the localized time dependent system-bath couplings. Two variants of chain mappings (Lanczos and block Lanczos) are presented in Sec.~\ref{sec:theory}. In Sec.~\ref{sec:application}, we apply both chain mappings to the singlet fission system, and calculate its time evolution.


\section{THEORY AND METHODS}
\label{sec:theory}
A general spin-Boson Hamiltonian is
\begin{align}
\begin{split}
    \hat{H}=&\Delta_x\hat{\sigma}_x+\hat{\sigma}_x\otimes\int{d\omega\,h_x(\omega) (\hat{a}_{\omega}^{\dagger}+\hat{a}_{\omega})}+\Delta_z\hat{\sigma}_z \\
            &+\hat{\sigma}_z\otimes\int{d\omega\,h_z(\omega) (\hat{a}_{\omega}^{\dagger}+\hat{a}_{\omega})}+\int{d\omega\,\omega \hat{a}_{\omega}^{\dagger}\hat{a}_{\omega}}.
\end{split}
\end{align}
$\hat{\sigma}_x$ and $\hat{\sigma}_z$ are the Pauli matrices, $\Delta_x$ and $\Delta_z$ are prefactors, $\hat{a}_\omega$ ($\hat{a}^\dagger_\omega$) is the annihilation (creation) operator for bosons with frequency $\omega$; $h_z(\omega)$ and $h_x(\omega)$ are the frequency-dependent couplings between the spin and the bath (through $\hat{\sigma}_x$ and $\hat{\sigma}_z$). We term them as $x$-interactions and $z$-interactions, respectively. The Hamiltonian is discretized as
\begin{align}
\begin{split}
    \hat{H}=&\Delta_x\hat{\sigma}_x+\underbrace{\hat{\sigma}_x\otimes \sum_{i=0}^N {\xi_{i} (\hat{a}_{i}^{\dagger}+\hat{a}_{i})}}_{x\text{-interactions}}\\
            &+\Delta_z\hat{\sigma}_z+\underbrace{\hat{\sigma}_z\otimes \sum_{i=0}^N {\zeta_{i} (\hat{a}_{i}^{\dagger}+\hat{a}_{i})}}_{_{z\text{-interactions}}} + \underbrace{\sum_{i=0}^N \omega_{i} \hat{a}_{i}^{\dagger}\hat{a}_{i}}_{\mathrm{bath}}.
\end{split}
\label{eq:dis-hamiltonian}
\end{align}
In the discretization, we use the Legendre polynomial discretization  \cite{de2015discretize} and choose the same discretization nodes (i.e., discrete frequencies) for the two spectral densities $J_{x} = \pi h^2_x(\omega)$ and $J_z(\omega) = \pi h^2_z(\omega)$. Any one of the two sets of discrete system-bath coupling coefficients $\{\xi_i\}$ and $\{\zeta_i\}$ can be used to generate a chain mapping, which allows the system-bath interaction terms characterized by this set of couplings to be transformed to a nearest-neighbor chain form  \cite{de2015discretize,chin2010exact,lacroix2021unveiling}. 
We refer to the chain mapping generated by $\{\xi_i\}$  as mapping $\mathbf{X}$, and the mapping generated by $\{\zeta_i\}$ as mapping $\mathbf{Z}$. It is generally impossible to transform both the $x$ and $z$-type system-bath interaction terms to the chain form simultaneously, unless the vectors $(\xi_1, \ldots,\xi_N)$ and $(\zeta_0,\ldots,\zeta_N)$ are parallel, so that the $\mathbf{X}$ and $\mathbf{Z}$ mappings are identical  \cite{dunnett2021influence}. The optimal nearest-neighbor chain form of the Hamiltonian in Eq.~\eqref{eq:dis-hamiltonian} therefore does not exist. Although one can simulate the open quantum system without transformation (i.e., the star form as in Eq.~\eqref{eq:dis-hamiltonian}, see details  in \footnote{Since the spin interacts with each boson, the couplings connect the spin to the bosons, forming a star pattern.}), the numerical cost is large  \cite{liu2021improved}. A strategy to circumvent this difficulty, and to enable efficient MPS simulations, is to transform the Hamiltonian of Eq.~\eqref{eq:dis-hamiltonian} into the interaction picture with respect to the bath, 
\begin{align}
\begin{split}
    \hat{H}_I=&\Delta_x\hat{\sigma}_x+\hat{\sigma}_x\otimes \sum_{j=0}^N{\xi_{j} (\hat{a}_{j}^{\dagger}e^{i\omega_j t}+\hat{a}_{j}e^{-i\omega_j t})} \\
        &+\Delta_z\hat{\sigma}_z +\hat{\sigma}_z\otimes \sum_{j=0}^N {\zeta_{j} (\hat{a}_{j}^{\dagger}e^{i\omega_j t}+\hat{a}_{j}e^{-i\omega_j t})},
\end{split}
\label{eq:int-hamiltonian}
\end{align}
and to use either of the two chain mappings ($\mathbf{X}$ or $\mathbf{Z}$) to transform the interaction-picture Hamiltonian.

\begin{figure}
	\includegraphics[width=\linewidth]{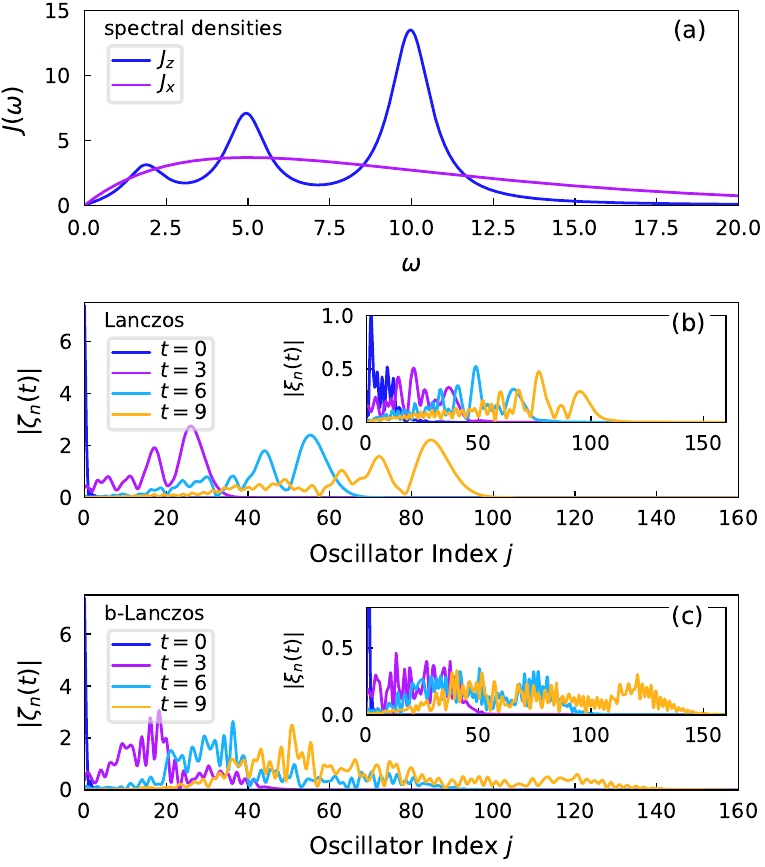}
	\caption{(a): The spectral densities $J_z$ and $J_x$ characterizing the $z$  and $x$ system-bath couplings. $J_z(\omega)=L(\omega, \Omega_1, \eta,\lambda) + L(\omega, \Omega_2, \eta,\lambda) + L(\omega, \Omega_3, \eta,\lambda)$ with $\Omega_1=2$, $\Omega_2=5$, $\Omega_3=10$, $\eta=1.5$, and $\lambda=1$. The Lorentzian function is  $L(\omega, \Omega, \eta,\lambda)=\frac{2\lambda \Omega^2 \eta \omega}{(\omega^2-\Omega^2)^2+\eta^2\omega^2}$.
    $J_x(\omega)=\lambda \omega e^{-\omega/\omega_c}$ with $\lambda=2$,  $\omega_c=5$. 
    (b): The absolute values of the Lanczos-transformed $z$-couplings $|\zeta_j(t)|$ and $x$-couplings $|\xi_j(t)|$ for bath oscillators (indexed by $n$, see Eq.~\eqref{eq:int-hamiltonian-chain-mapped} with the definitions of $\xi_j(t)$ and $\zeta_j(t)$). The Lanczos transformation is generated by the $z$-couplings $\{\zeta_i\}$ [Eq.~\eqref{eq:dis-hamiltonian}]. 
    (c): The absolute values of the block Lanczos transformed $z$ couplings $|\zeta_j(t)|$ and $x$-couplings $|\xi_j(t)|$ for bath oscillators (see Eq.~\eqref{eq:int-hamiltonian-chain-mapped} and Eq.~\eqref{eq:block-lanczos-definition}).
    }
	\label{fig:wave}
\end{figure}

Mapping $\mathbf{X}$, when applied to the $x$-coupling terms, can produce a set of time-dependent couplings that resemble the desired chain form, as shown in Ref~ \cite{liu2021improved}. For the $z$-couplings, the chain mapping $\mathbf{X}$ cannot transform the $z$-interaction terms into a nearest-neighbor interacting chain form. However, in the interaction picture, mapping $\mathbf{X}$ produces time dependent $z$-type couplings with the localized traveling-wave pattern (see Fig.~\ref{fig:wave}), which is essential for efficiently simulating the time-dependent Hamiltonian Eq.~\eqref{eq:dis-hamiltonian} (see Ref~\cite{liu2021improved}).

Chain mapping $\mathbf{X}$ can be generated as follows. We transform $\mathbf{w} = \operatorname{diag}(\omega_0,\cdots,\omega_N)$ into a tridiagonal chain form by applying an orthogonal transformation to $\mathbf{w}$: $\mathbf{P}^T\mathbf{w}\mathbf{P} =\mathbf{W}$. The orthogonal matrix $\mathbf{P}$ is then used to generate a new set of bath modes: $\hat{b}^\dagger_j = \sum_i P_{ij} \hat{a}^\dagger_i$ (and $\hat{b}_j = \sum_i P_{ij} \hat{a}_i$). The bath Hamiltonian can be written in terms of the new operators $\hat{b}^\dagger_i$ and $\hat{b}_i$ as
\begin{align}
    H_b = \sum_{i}{\omega_{i} \hat{a}_{i}^{\dagger}\hat{a}_{i}} =
    \mathbf{b^\dagger} \mathbf{W} \mathbf{b}
    \label{eq:harmoinc-bath-hamiltonian}
\end{align}
where $\mathbf{\hat{b}^\dagger}=(\hat{b}_0^\dagger, \ldots, \hat{b}_N^\dagger)$ and $\mathbf{b}=(\hat{b}_0, \ldots, \hat{b}_N)^T$ represent the new bath modes, and the tridiagonal matrix $\mathbf{W}$ is
\begin{align}
    \mathbf{W}=\begin{pmatrix}
    \alpha_0 & \beta_1 &  &  & \\
    \beta_1 & \alpha_1 & \beta_2 & & \\
     & \ddots & \ddots & \ddots & \\
    \ & & \beta_{N-1} & \alpha_{N-1} & \beta_N\\
     &  &  & \beta_N & \alpha_N
    \end{pmatrix}.
\end{align}
By further requiring the $x$-interaction terms to become the interactions between only the system and the zeroth mode: $\hat{\sigma}_x\otimes\sum_i\xi_i(\hat{a}^\dagger_i+\hat{a}_i)\to \hat{\sigma}_x\otimes\tilde{\xi}_0(\hat{b}^\dagger_0+\hat{b}_0)$,
with $\tilde{\xi}_0=\sqrt{\sum_i\xi_i^2}$, 
we obtain the first column of $\mathbf{P}$: $P_{0i} = \xi_i/\tilde{\xi}_0$  \cite{cederbaum2005short,tamascelli2019efficient,chin2010exact,prior2010efficient}. The orthogonal matrix $\mathbf{P}$ is then determined by solving 
\begin{align}
	\mathbf{wP} = \mathbf{PW}.
\end{align}
for the other columns of $\mathbf{P}$.

The orthogonal matrix $\mathbf{P}$ produced by the above procedure  links the old operators $(\hat{a}^\dagger_i,\hat{a}_i)$ to the new ones $(\hat{b}^\dagger_j,\hat{b}_j)$. One can replace the old operators in Eq.~\eqref{eq:int-hamiltonian} with the new operators and obtain a set of time-dependent couplings:
\begin{align}
\begin{split}
    \hat{H}_I = & \Delta_x \hat{\sigma}_x + \hat{\sigma}_x\otimes\sum_{jk=0}^N P_{jk}{\xi_{j} (\hat{b}_{k}^{\dagger}e^{i\omega_j t}+\hat{b}_{k}e^{-i\omega_j t})} \\
                &+\Delta_z\hat{\sigma}_z +\hat{\sigma}_z\otimes\sum_{jk=0}^N P_{jk}{ \zeta_j(\hat{b}_{k}^{\dagger}e^{i\omega_j t}+ \hat{b}_{k}e^{j\omega_i t})} \\
       =&\Delta_x\hat{\sigma}_x+\hat{\sigma}_x\otimes\sum_{k=0}^N  [\xi_{k}(t)\hat{b}_{k}^{\dagger}+\xi^*_{k}(t)\hat{b}_{k}] \\
        &+\Delta_z\hat{\sigma}_z +\hat{\sigma}_z\otimes\sum_{k=0}^N[ \zeta_{k}(t)\hat{b}_{k}^{\dagger} + \zeta^*_{k}(t)\hat{b}_{k} ]
\end{split}
\label{eq:int-hamiltonian-chain-mapped}
\end{align}
where $\xi_{k}(t)=\sum_{j} P_{jk}\xi_j e^{-i\omega_j t}$ and $\zeta_{k}(t)=\sum_jP_{jk}\zeta_j e^{-i\omega_j t}$. 
$\mathbf{P}$ is not the only possible orthogonal transformation that can be used to transform the bath Hamiltonian of Eq.~\eqref{eq:harmoinc-bath-hamiltonian}; rather, any orthogonal matrix can be used to transform the diagonal harmonic bath Hamiltonian and may be used in Eq.~\eqref{eq:dis-hamiltonian} to obtain a new interaction-picture Hamiltonian. However, different orthogonal transformations produce different numerical efficiencies. As shown in  \cite{liu2021improved}, the numerical efficiency of simulating the time-dependent Hamiltonians of Eq.~\eqref{eq:int-hamiltonian-chain-mapped} depends on the properties of the orthogonal matrix ($\mathbf{P}$ or other possible orthogonal matrices).

\begin{figure}
    \centering
    \includegraphics[width=\linewidth]{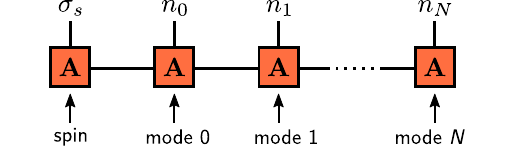}
    \caption{The MPS of the spin and bath in Eq.~\eqref{eq:int-hamiltonian-chain-mapped}: $\ket{\psi_{sb}}=\sum_{\sigma_s n_0\cdots n_N}\mathbf{A}^{\sigma_s}\mathbf{A}^{n_0}\cdots\mathbf{A}^{n_N}$. $\sigma_s$ is the spin basis index. $n_0,\ldots,n_N$ are indices of bath mode eigenbasis. The left-most matrix $\textbf{A}^{\sigma_s}$ corresponds to the spin, while $\textbf{A}^{n_0},\ldots,\textbf{A}^{n_N}$ correspond to bath modes 0 to $N$, respectively.
    }
    \label{fig:MPS}
\end{figure}

The Hamiltonian [Eq.~\eqref{eq:int-hamiltonian-chain-mapped}] has a natural ordering of sites for the MPS representation of wave functions. The first site of the MPS is the spin; the others represent bath oscillators, and the oscillators with smaller indices are closer to the spin site, as shown in Fig.~\ref{fig:MPS}.

Any orthogonal matrix can replace the orthogonal matrix $\mathbf{P}$ without changing the dynamics of the spin, as noted above. However, the numerical efficiency of a MPS simulation depends crucially on the choice of the orthogonal matrix.  Here, we provide an alternative to $\mathbf{P}$. We construct an orthogonal matrix with its first two columns orthogonal to the vector space spanned by $\{\zeta_i\}$ and $\{\xi_i\}$ in Eq.~\eqref{eq:dis-hamiltonian}  \cite{gindensperger2006short,cederbaum2005short}. One can require the orthogonal matrix (denoted $\mathbf{Q}$) to transform the diagonal bath Hamiltonian into a chain of oscillators with nearest and next-nearest-neighbor couplings so that the desired traveling wave structure still appears, thus reducing the entanglement growth over the course of time. The orthogonal matrix $\mathbf{Q}$ is then uniquely determined by solving the following linear equation recursively  \cite{giantAtom,blockLanczosAlgorithm}:
\begin{align}
\begin{split}
    & \mathbf{Q}^T\mathbf{w}\mathbf{Q} = \\
    & \begin{pmatrix}
        \alpha_0 & \beta_1 & \kappa_2 & 0 &  &  &  \\
        \beta_1  & \alpha_1 & \beta_2 & \kappa_3 & 0 &  & \\
        \kappa_2 & \beta_2  & \alpha_2 & \beta_3 & \kappa_4 & 0 &  &      \\
        0        & \kappa_3 & \beta_3  & \alpha_3 & \beta_4 & \kappa_5 & 0 \\
        & \ddots & \ddots & \ddots & \ddots & \ddots & \ddots & \ddots \\
        & & 0 & \kappa_{N-3} & \beta_{N-3} & \alpha_{N-3} & \beta_{N-2} & \kappa_{N-1} & 0 \\
        & & & 0 & \kappa_{N-2} & \beta_{N-2} & \alpha_{N-2} & \beta_{N-1} & \kappa_{N} \\
        & & & &  0 & \kappa_{N-1} & \beta_{N-1} & \alpha_{N-1}& \beta_N \\
        & & & & & 0 & \kappa_{N} & \beta_N & \alpha_N
    \end{pmatrix}.
\end{split}
\label{eq:block-lanczos-definition}
\end{align}
The orthogonal matrix $\mathbf{Q}$ can replace $\mathbf{P}$ (the Lanczos matrix) in Eq.~\eqref{eq:int-hamiltonian-chain-mapped}. The matrix $\mathbf{Q}$ is known as the block Lanczos matrix  \cite{blockLanczosAlgorithm}.

To show that a single transformation ($\mathbf{P}$ or $\mathbf{Q}$) can generate two sets of couplings with a traveling-wave pattern  \cite{liu2021improved}, we use two distinct spectral densities
for $x$ and $z$-couplings to obtain the discretized Hamiltonian [Eq.~\eqref{eq:dis-hamiltonian}]. The discretized Hamiltonian is then transformed by the orthogonal transformation generated by the Lanczos method regarding either one of the two couplings [Fig.~\ref{fig:wave}, panel (b)], or, instead, by the transformation $\mathbf{Q}$ generated by the block Lanczos method using two spectral densities simultaneously [Fig.~\ref{fig:wave}, panel (c)]. The parameters of the spectral densities are noted in the captions of Fig.~\ref{fig:wave}.
The traveling-wave pattern appears in the time-dependent couplings ($\xi_i(t)$ and $\zeta_i(t)$) of both the $x$ and $z$-interactions. This pattern ensures the efficiency of the interaction-picture MPS simulations for the chain-mapped Hamiltonian with two distinct spectral densities, since the interactions between the system and the bath are localized spatially to a relatively narrow window of bath oscillators   \cite{liu2021improved}.

\section{Application}
\label{sec:application}
The spin-boson model is widely used to describe transitions between molecular electronic states coupled to nuclear vibrations, environmental phonons, and photon modes of the radiation field  \cite{10.1093/oso/9780198529798.003.0018}. This approach is frequently used to simulate non-adiabatic dynamics, including bridge-mediated electron transfer  \cite{doi:10.1063/1.4950888, doi:10.1063/1.4990739, doi:10.1063/5.0027976}.

To explore the validity of the method developed in Sec.\ref{sec:theory}, we apply the method to a spin-boson system with both diagonal and non-diagonal couplings. In organic crystals where intramolecular and intermolecular vibrations induce diagonal and off-diagonal exciton-phonon couplings  \cite{Berkelbach1, Berkelbach2}, two spectral densities are required to describe the coupling. We model singlet fission, where a singlet exciton state $\ket{S_1}$ evolves into two triplet excitons $\ket{TT}$  \cite{SmithSinglet}. In some organic materials, singlet fission is facilitated by an intermediate charge transfer state that couples the singlet and triplets states  \cite{HuangSFmodel}. The rate of singlet fission is influenced by both diagonal and off-diagonal vibronic interactions. The Hamiltonian is represented by Eq.~\eqref{eq:dis-hamiltonian}, including the Hamiltonian of the exciton states, harmonic nuclear vibrations, and the interactions between them. Here, we use the model singlet fission Hamiltonian of Huang et al.  \cite{HuangSFmodel}:
\begin{align*}
\begin{split}
    \hat{H} = & \Delta_x (\ket{S_1}\bra{TT}+h.c.) + \Delta_z \ket{S_1}\bra{S_1} +  \sum_{i} \hbar \omega_i \hat{a}^\dagger_i\hat{a}_i + \hat{H}_{sb}, \\
    \hat{H}_{sb} = & \big(\sqrt{\lambda_{\smash{S_1}}}\ket{S_1}\bra{S_1} + \sqrt{\lambda_{\smash{TT}}}\ket{TT}\bra{TT} \big)\otimes \sum_{i}\zeta_i (\hat{a}^\dagger_i + \hat{a}_i)  \\
    & + \sqrt{\lambda_{\smash{o.d.}}}\big(\ket{S_1}\bra{TT} + h.c.\big) \otimes \sum_{i}\xi_i (\hat{a}^\dagger_i + \hat{a}_i).
\end{split}
\end{align*}
$\lambda$ values characterize the magnitude of system-bath couplings.
Two Lorentzian spectral densities $J_z,\,J_x$ define the diagonal and off-diagonal couplings, $\{\xi_i\}$ and $\{\zeta_i\}$, respectively:
\begin{align}
    J_z(\omega) & = \pi\sum_{i}\zeta_i^2\delta(\omega-\omega_i) = \frac{4\gamma_{diag}\omega_{diag}^2\omega}{(\omega^2-\omega_{diag}^2)^2+4\gamma_{diag}^2\omega^2},\\
    J_x(\omega)&=\pi\sum_{i}\xi_i^2\delta(\omega-\omega_i) = \frac{4\gamma_{o.d.}\omega_{o.d.}^2\omega}{(\omega^2-\omega_{o.d.}^2)^2+4\gamma_{o.d.}^2\omega^2}.
\end{align}
$\omega_{diag}$ and $\omega_{o.d.}$ are vibrational frequency centers of the Lorentzian spectral densities, and $\gamma_{diag}=\gamma_{o.d.}=1\,\mathrm{ps}^{-1}$ are vibrational relaxation rates. Values of the other parameters appear in Table~\ref{table:parameter}.
\begin{table}[!h]
    \centering
    \begin{tabular}{ll}
        \toprule
         Quantity &  Value \\ 
         \colrule
         Energy difference $\Delta_z$ &  $100\,\mathrm{meV}$ \\
         Electronic coupling $\Delta_x$ & $20\,\mathrm{meV}$ \\
         Coupling strength $\lambda_{S_1}$  & $0.7\hbar\omega_{diag}$ \\
         Coupling strength $\lambda_{S_{TT}}$  & $1.4\hbar\omega_{diag}$ \\
         Coupling strength $\lambda_{o.d.}$ & $0.1 \hbar\omega_{o.d.}$ \\
         Frequency $\omega_{diag}$, $\omega_{o.d.}$ & See Fig.~\ref{fig:population}\\
         Spectral density cutoff & [0, $800\,\mathrm{cm}^{-1}$] \\
         Bath temperature & $0\,\mathrm{K}$ \\
         Number of bath oscillators & $300$ \\
         SVD cutoff for MPS & $10^{-4}$ \\
         Energy levels of bath modes & 160 \\
         \botrule
    \end{tabular}
    \caption{Parameters used in the singlet fission simulations.}
    \label{table:parameter}
\end{table}

\begin{figure}[!h]
    \centering
    \includegraphics[width=\linewidth]{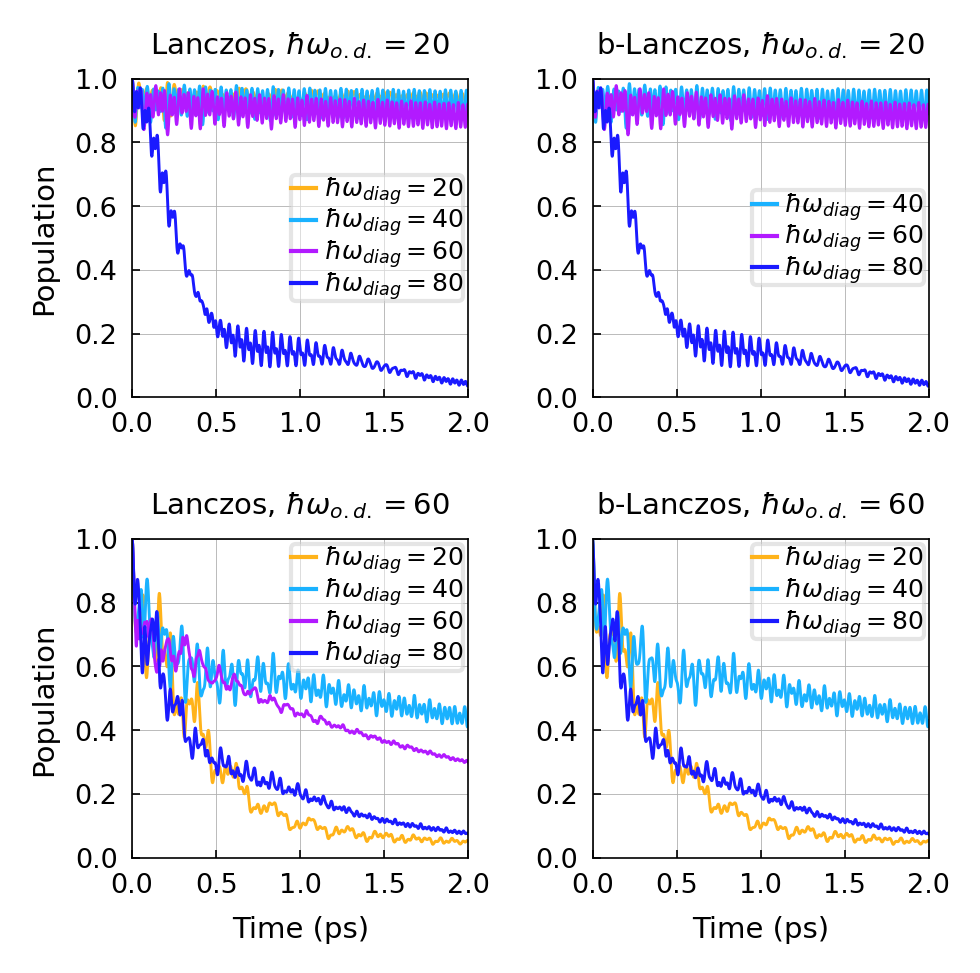}
    \caption{Time evolution of the singlet state population. The $x$-axis shows the time in picosecond, and the $y$-axis shows the population of the singlet state. The left column shows simulation results for the Lanczos transformation, while the right column presents population evolution of block Lanczos transformation. Each panel shows the simulation results for diagonal vibrational frequencies of $20,\,40,\,60,\,80\,\mathrm{meV}$. For the block Lanczos method, $\omega_{diag}$ and $\omega_{o.d.}$ must have distinct values, otherwise the matrix $\mathbf{Q}$ in Eq.~\eqref{eq:block-lanczos-definition} would not become orthogonal.}
    \label{fig:population}
\end{figure}

We ran the singlet fission simulation with an initial state $\ket{\varphi(0)}=\ket{S_1}\ket{0}$, where $\ket{0}$ indicates that all bath harmonic oscillators are in their ground state at $T=0\,\mathrm{K}$ and $t=0$. Varying the parameters $\omega_{diag},\, \omega_{o.d.}$ in $J_z$ and $J_x$, we compute the population evolution of the singlet state for different central mode frequencies ($\omega_{diag}$ and $\omega_{o.d.}$) of  intermolecular and intramolecular vibrations, as shown in Fig.~\ref{fig:population}. Our Lanczos and block Lanczos simulation results are consistent in the time evolution of singlet state population, and the time evolution behavior is similar to previous findings for the singlet fission system  \cite{HuangSFmodel}. However, we find different singlet state populations after $1\,\mathrm{ps}$, which continues to drop in our simulation but oscillates around $0.4$ in the earlier simulations. The discrepancy may be due to the inadequate multiplicity in their multi-$D_2$ ansatz  \cite{HuangSFmodel}.

In the upper panels of Fig.\ref{fig:population}, where the off-diagonal coupling is weak compared to the diagonal coupling, the singlet fission dynamics is dominated by diagonal vibration modes and produces a resonant transition when $\hbar\omega_{diag}=80\,\mathrm{meV}$. However, when $\hbar\omega_{o.d.}$ is increased to $60\,\mathrm{meV}$, the off-diagonal coupling begins to break the strict condition of resonant transitions in the singlet fission dynamics and slows the transition rate accordingly.

\begin{figure}
    \centering
    \includegraphics[width=\linewidth]{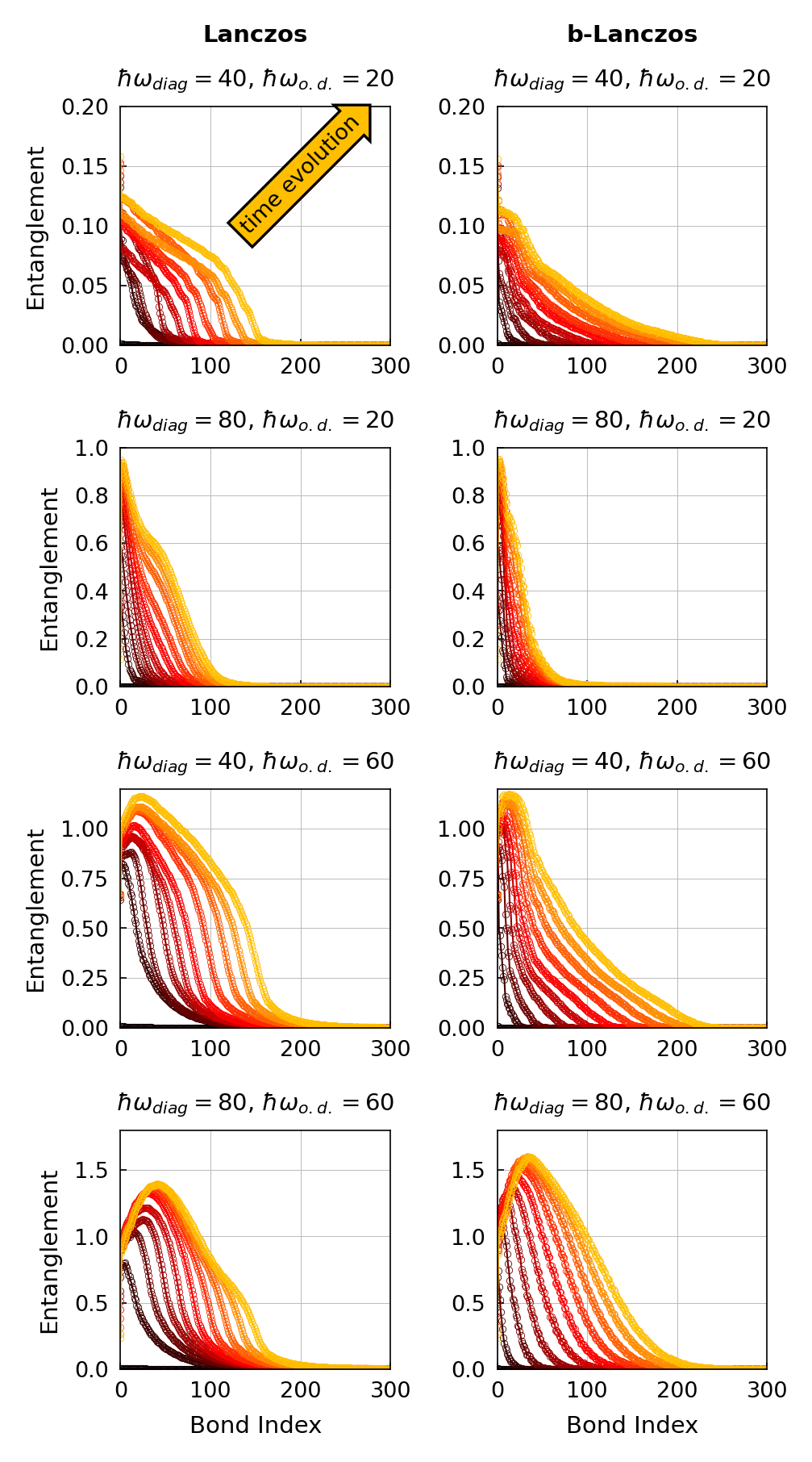}
    \caption{Entanglement growth of MPS bonds (i.e, bipartite divisions) with time. The evolution is computed using the Lanczos and block Lanczos transformed Hamiltonians [Eq.~\eqref{eq:int-hamiltonian-chain-mapped}]. The y-axes of the panels are entanglement. The time evolution direction is labeled by the arrow in the first panel. A bond index of $0$ represents the bond between the spin and first bath mode, while other indices represent bonds between bath modes. The block Lanczos method shows higher values of entanglement (see Sec.\ref{sec:application} for the definition), especially for the $\hbar\omega_{diag}=80\,\mathrm{meV},\,\hbar\omega_{o.d.}=60\,\mathrm{meV}$ panels. These results demonstrate that MPS computations using the Lanczos transformation have a faster simulation speed than the block Lanczos transformation.}
    \label{fig:entanglement}
\end{figure}

In a MPS, a quantum system is divided into two parts, with entanglement between them. The division is referred to as a bond, which introduces the concept of entanglement between the two parts.
The entanglement for the division (bond) measures how many states in one of the two parts are needed to represent the wave function faithfully. As such, the larger the entanglement, the slower the simulation becomes.
Fig.~\ref{fig:entanglement} shows the entanglement growth of MPS bonds with time evolution, providing comparisons between Lanczos and block Lanczos algorithms. 
The growth of the MPS entanglement behaves slightly differently for the two methods. However, the overall time-evolution patterns of entanglement are similar for the Lanczos and block Lanczos methods, showing a propagation of entanglement away from the spin.
We conclude from Fig.~\ref{fig:entanglement} that the Lanczos method provides a faster simulation speed than  is found using the block Lanczos algorithm. This is understood because the entanglement is generated not only  by nearest-neighbor interactions but also by next-nearest-neighbor interactions.

\section{Conclusions}
\label{sec:conclusion}
We developed a DMRG-based approach to simulate multi-channel open quantum system with two distinct spectral densities.
We discretized the spectral densities of the system-bath couplings using the Legendre polynomial discretization method, producing two sets of coupling strengths $\{\xi_i\}$ and $\{\zeta_i\}$. Then, we transformed the Hamiltonian to the interaction picture, and applied an orthogonal matrix to perform chain mapping and to generate new sets of modes. We developed two kinds of orthogonal matrices, with one transforming the diagonal vibrational frequencies into a tridiagonal matrix (generated by the Lanczos algorithm), and the other resulting in a block tridiagonal matrix (generated by the block Lanczos algorithm). The Lanczos algorithm, producing the chain mapping from either $\{\xi_i\}$ or $\{\zeta_i\}$, generates only nearest-neighbor interactions between chain modes. The block Lanczos method, in contrast, generates additional next-nearest-neighbor interactions, and produces the orthogonal matrix taking both $\{\xi_i\}$ and $\{\zeta_i\}$ into account. In the interaction-picture chain mapping scenario, the interactions between the system and bath are limited to a localized space within the MPS chain, substantially accelerating the simulation.

We applied the two chain mapping techniques to simulate singlet fission, where the singlet and triplet states are coupled non-adiabatically to a bath through two Lorenztian spectral densities. By varying the central vibronic frequencies of the spectral densities, we found different singlet fission rates that characterize the decay rate of the singlet state population. The singlet fission findings are in qualitative agreement with previous results but differ at  longer times. We also analyzed the efficiencies of the Lanczos and block Lanczos methods by examining the entanglement evolution of the MPS bonds. We found that the block Lanczos method generally has a higher entanglement value than the Lanczos approach, leading to the higher efficiency for the Lanczos method. The lower efficiency of the block Lanczos approach is attributed to the introduction of next-nearest-neighbor interactions, since next-nearest-neighbor interactions entangle a bath mode with many  other bath modes.

\begin{acknowledgments}
	This material is based upon work supported by the National Science Foundation under Grant No. 1955138.
\end{acknowledgments}

\bibliography{references}

\end{document}